\documentclass[aps,prl,twocolumn,superscriptaddress,showpacs]{revtex4}

\usepackage{graphicx}% Include figure files
\usepackage{dcolumn}% Align table columns on decimal point
\usepackage{bm}% bold math

\def\qsgw{\mbox{QS$GW$}}
\def\qsgwff{\mbox{QS$GW$[$f$-$f$]}}
\def\efermi{$E_{\rm F}$}

\def\ldau{LDA+$U$}
\def\ff{ff}

\def\spdsubsystem{\mbox{$spd$ subsystem}}
\def\fsubsystem{\mbox{$f$ subsystem}}

\def\GWLDA{${G^{\rm LDA}}{W^{\rm LDA}}$}
\def\fup{$f^\uparrow$}
\def\fdn{$f^\downarrow$}

\begin{document}

\title{$GW$ method applied to localized $4f$ electron systems}

\author{Athanasios N. Chantis, Mark van Schilfgaarde and Takao Kotani}
\affiliation{School of Materials, Arizona State University, Tempe, Arizona, 85287-6006, USA}

\date{\today}

\begin{abstract}

We apply a recently developed quasiparticle self-consistent $GW$ method
(\qsgw) to Gd, GdAs, GdN and ErAs. We show that \qsgw\ combines advantages
separately found in conventional $GW$ and \ldau\ theory, in a simple and
fully \emph{ab initio} way.  \qsgw\ reproduces the experimental occupied
$4f$ levels well, though unoccupied levels are systematically
overestimated.  Properties of the Fermi surface responsible for electronic
properties are in good agreement with available experimental data.  GdN is
predicted to be very near a critical point of a first-order metal-insulator
transition.

%We show that a self-consistent form of the $GW$ method can successfully
%describe the electronic structure of materials that have both moderately
%correlated $spd$ and strongly correlated $f$ electrons.  Self-consistency
%is essential.  We predict a new effect, a first order metal-insulator
%transition in GdN, driven by dielectric function changes.  Because the
%method is rigorously grounded in many-body perturbation theory without any
%special ansatz, it can be used to examine approximations in some standard
%semiempirical approaches.

\end{abstract}

\pacs{71.15-m,71.10-w,71.20-Eh}

\maketitle

% ----------------- Start ---------------------
Most 4$f$ compounds belong to a class of materials whose electronic
structure can be approximately described in terms of the coexistence of two
subsystems --- a localized \fsubsystem, and an itinerant \spdsubsystem.
States near the Fermi energy \efermi\ predominantly consist of the latter;
4$f$ electrons largely play a passive role except to spin-polarize the
\spdsubsystem\ through an indirect exchange mechanism.  Describing both
subsystems in the framework of \emph{ab initio} electronic structure
methods, however, poses a rather formidable challenge.  The most widely
used method, the local density approximation (LDA), has been an immensely
successful tool that reasonably predicts ground-state properties of weakly
correlated systems.  The LDA is much less successful at predicting optical
properties of such systems, and its failures become serious when
correlations become moderately strong.
%ground state properties are often poorly described\,\cite{filippetti03}.
It fails catastrophically for open $f$-shell systems, leaving $f$ electrons
at \efermi.  To surmount these failures, a variety of strategies to
extend the LDA have been developed.  These include exact exchange (EXX)
\cite{Kotani98}, self-interaction-correction\,\cite{aerts} (SIC),
\ldau\,\cite{dowben,komesu03,petukhov-eras-gas}, and more recently LDA+DMFT
(dynamical mean-field theory)\,\cite{savrasov06}.  As a consequence, they
have serious problems, both formal and practical.  SIC, LDA+$U$ and
LDA+DMFT add nonlocal potentials to certain localized electrons in a
special manner, leaving some ambiguity about how a localized electron state
is defined, and how the double-counting term should be subtracted.
Further, such approaches are specialized; they cannot remedy the LDA's
inadequate description of itinerant \spdsubsystem{}s (e.g. its well-known
underestimate of semiconductor bandgaps), which is the relevant one for
transport properties.  Thus, they are problematic for $4f$ materials such
as the rare earth monopnictides we study here.
%EXX has less ambiguity, but suffers from many other failures.  When RPA
%correlation is added to the EXX potential, essentially the LDA result is
%recovered in Si\,\cite{Kotani98}.
%The (improvement in EXX semiconductor bandgaps is artifact of fortituous
%cancellation between the discontinuity in the exchange potential and
%neglect of correlation).
%Moreover, local potentials cannot break time-reversal symmetry---an
%essential feature in rare earth elements\,\cite{walter} and CoO.  
The standard $GW$ (i.e. 1-shot $GW$ as perturbation to the LDA, or \GWLDA)
significantly improves on the LDA's description of itinerant
\spdsubsystem{}s, but it has many shortcomings\,\cite{mark06adeq}; and it
fails qualitatively in open $f$ systems, in much the same way as the LDA
fails~\cite{mark06adeq}.

In short, the present status consists of an unsatisfactory patchwork
%of \emph{ab initio} and quasi-\emph{ab initio} methods, each with successes in
of methods, each with successes in improving some property in one or
another class of materials.  Here we show, for the first time, that a
recently developed quasiparticle self-consistent $GW$
method\,\cite{faleev,Mark,chantis,qpmethod} (\qsgw) can reliably describe
open $f$-shell systems, including the \spdsubsystem.  The $GW$ approximation is a
prescription for mapping the non-interacting Green function to the dressed
one, $G^0 \to G$.  Formally, $G$ can be calculated from any $G^0$. \qsgw\
gives a prescription to specify a (nearly) optimal mapping $G \to G^0$, so
that $G^0 \to G \to G^0 \to ...$ can be iterated to self-consistency.  At
self-consistency the quasiparticle energies of $G^0$ coincide with those of
$G$.  Thus \qsgw\ is a self-consistent perturbation theory, where the
self-consistency condition is constructed to minimize the size of the
perturbation.  \qsgw\ is parameter-free, independent of basis set and of
the LDA\,\cite{qpmethod}.  It contains \ldau\ kinds of effects, but no
subsystem is singled out for specialized treatment; there are no
ambiguities in double-counting terms, or in what is included and what is
left out of the theory.  We showed that \qsgw\ reliably describes a wide
range of $spd$ systems~\cite{Mark}.  Its success in describing $f$ systems
is important because it is not known whether the $GW$ method can reasonably
describe correlated $f$ electrons at all.
%We will show that the $f$ and $spd$ subsystems are both well described in
%all the RE systems we studied.

%The success with the compounds studied shows
%that $\qsgw$ shows much promise as a \emph{universal} approach that is
%suitable for any compound in the periodic table.  It is, to our knowledge,
%the only truly \emph{ab initio} method capable of doing so in a consistent,
%unified framework.

In addition, this work brings out some new fundamental points.  First, the
position of unoccupied $f$ levels is systematically overestimated.  
%We associate this error mainly with vertex corrections in the dielectric function
%(probably ladder diagrams). 
Second, we predict that GdN is on the cusp of
a new kind of first-order metal-insulator transition (MIT).  Last, we show
that the shifts in $f$ levels that \qsgw\ determines relative to LDA, lie
outside the degrees of freedom inherent in the standard \ldau\ method.  
The method enables us to reconstruct parameters used in the \ldau\
theory, as we will show.
%We show that we can use a subset of the \qsgw\ potential that mimics \ldau,
%which enables us to in principle reconstruct optimum choices for parameters
%such as $U$ and $J$ in \ldau.

%\qsgw\ was introduced in Refs.\,\cite{Mark} and \cite{faleev}.
Ref.\,\cite{qpmethod} gives some formal justification as to why \qsgw\
should be preferred to conventional self-consistent $GW$.  The orbital
basis
%(a variant of the full-potential linear muffin-tin orbital method)
and our development of the all-electron $GW$ are described in
Ref.\,\cite{mark06adeq}.  Local orbitals (e.g. $5f$ states) are essential
for reliable description of these systems in \qsgw.  It is also important
not to assume time-reversal symmetry in the open $f$ systems
\cite{LarsonREN}.

\begin{figure}[htbp]
\includegraphics[angle=0,width=0.23\textwidth,clip]{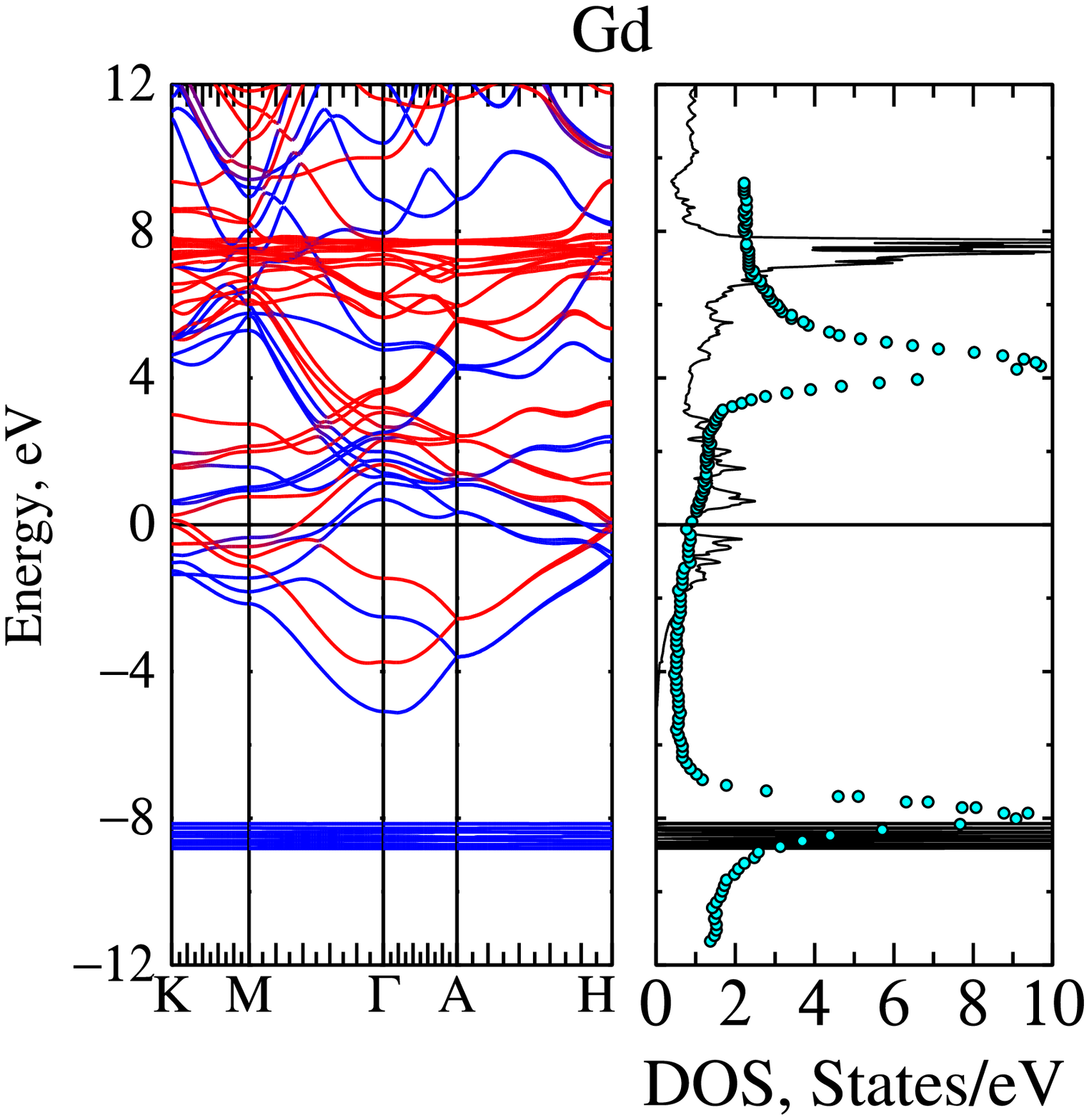}
\includegraphics[angle=0,width=0.23\textwidth,clip]{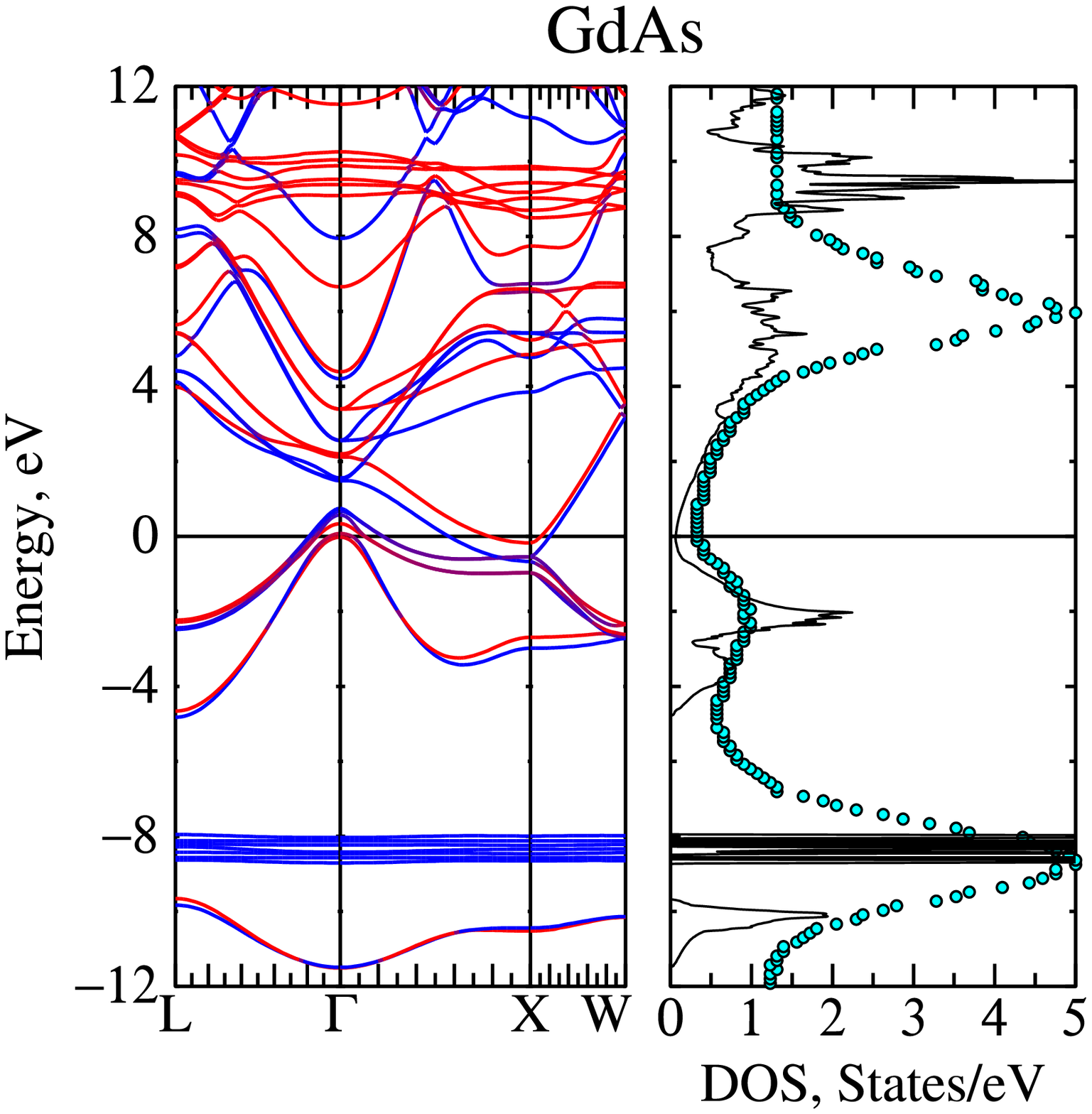}
\includegraphics[angle=0,width=0.23\textwidth,clip]{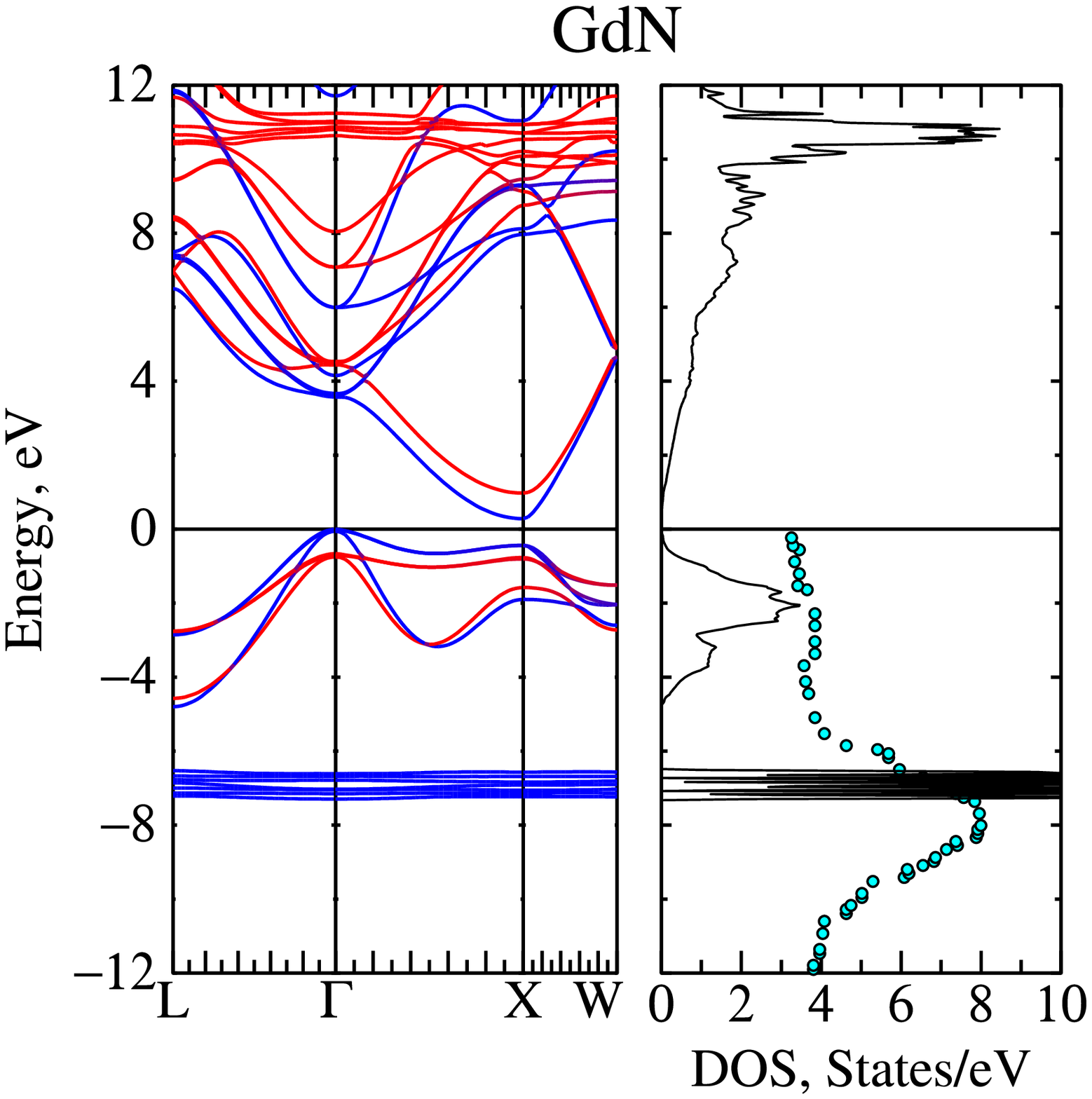}
\includegraphics[angle=0,width=0.23\textwidth,clip]{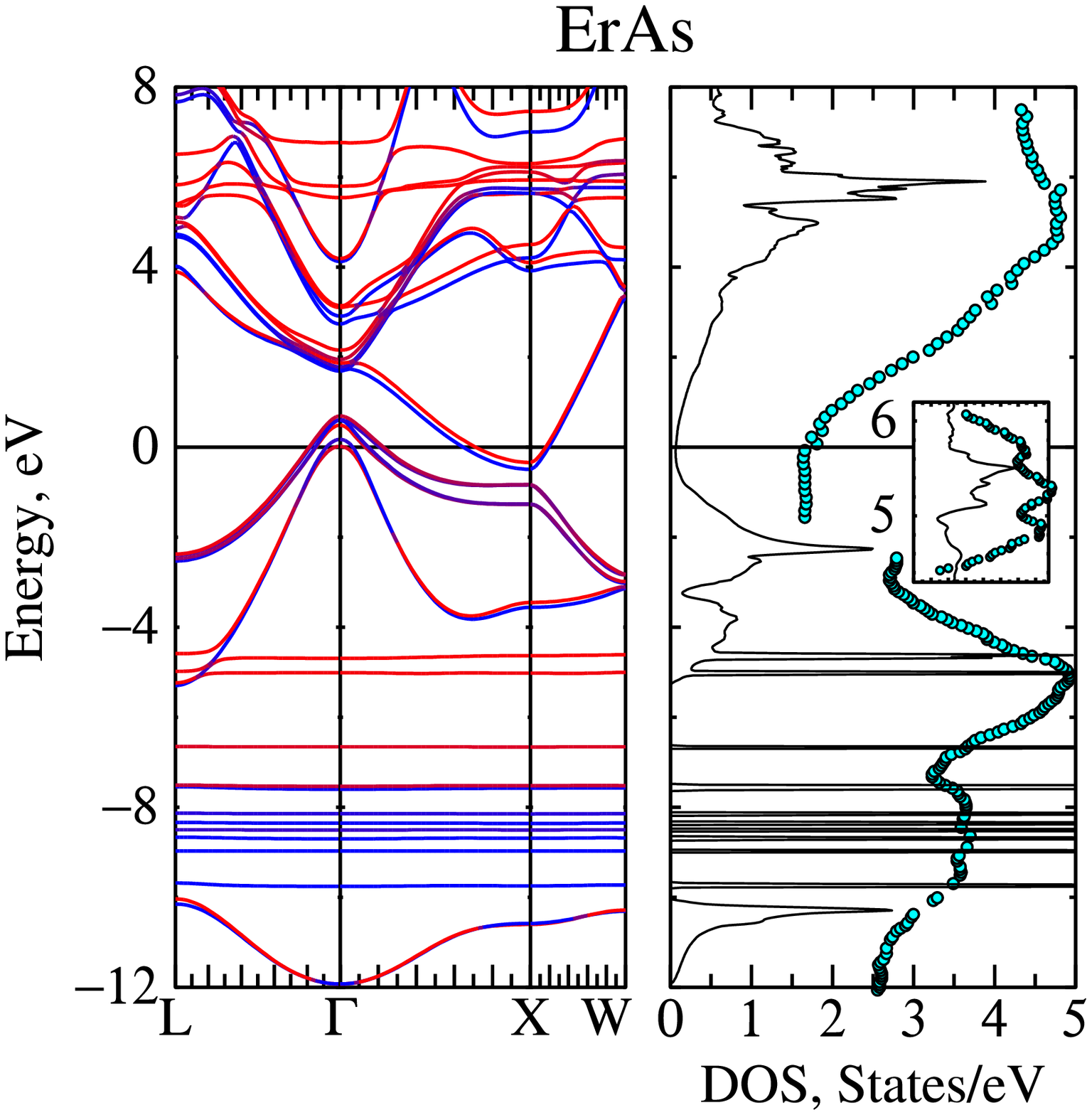}

\caption{ \small The \qsgw\ energy band structure of Gd, GdAs, GdN, and ErAs.
Right panels show DOS, together with experimental XPS and BIS data\cite{lang,yamada,leuenberger,komesu03} (circles).
\efermi=0\,eV.  Colors indicate the spin character
of the band (blue for majority and red for minority).
Lattice constants were taken to be 5.00\AA (GdN), 5.80\AA (GdAs), 5.73\AA (ErAs)
in the NaCl structure, and 3.64\AA (Gd) in hcp.
}
\label{fig:band}
\end{figure}

We considered the following $4f$ systems: Gd, Er, EuN, GdN, ErAs, YbN, and
GdAs.  Gd and Er are metals, while the rest are narrow-gap insulators or
semimetals.  \qsgw\ always shifts $4f$ levels away from \efermi.  The
electronic structure around \efermi\ is dominated by $spd$ electrons, which
we will consider later.  Fig.\,\ref{fig:band} shows energy bands and QP
density of states (DOS), for some cases, together with XPS (X-ray
photoemission) and BIS (bremsstrahlung isochromat) spectroscopies.

\begin{table}[htbp]
\caption{\qsgw\ spin and orbital moments (bohr), average position of $4f$ levels
relative to \efermi\ (eV), and corresponding peaks in XPS and BIS data (where available).
When \fdn\ or \fup\ states are split between 
occupied and unoccupied levels, average positions for both occupied and unoccupied
are given (top and bottom numbers).
When the occupied or unoccupied part of \fdn\ or \fup\ levels consist of multiple states split about the average
(see, e.g. ErAs bands in Fig.\,\ref{fig:band}), the range of splitting is denoted in
parentheses. 
}
\begin{center}
\begin{tabular}{|l|ccc|l|l|l|l|}
\hline
& 
\emph{$\mu $}$_{\rm spin}$ & 
\emph{$\mu $}$_{\rm orb}$& 
\emph{$\mu $}$_{\rm exp}$& 
\ \fup& 
XPS& 
\ \fdn& 
BIS \\
\hline
Gd& 
7.8& 
$-$& 7.6\footnotemark[1]&
-8.5& 
-8.0& 
\ 7.7& 
4.2 \\
\hline
& & & &
-8.1(2)& 
-8.4,-4.6
& 
-4(1)
& 
\\
\raisebox{1.5ex}[0pt]{Er}&
\raisebox{1.5ex}[0pt]{3.5}& 
\raisebox{1.5ex}[0pt]{6.0}& 
\raisebox{1.5ex}[0pt]{9.1\footnotemark[2]}& 
& &\ 5.1&2.1\\
\hline
& & & &
-6(2)& 
&& 
 \\
\raisebox{1.5ex}[0pt]{EuN}&
\raisebox{1.5ex}[0pt]{6.0}& 
\raisebox{1.5ex}[0pt]{\hspace{-4pt}-2.8}& &
\ 3.1&&
\ 9& 
 \\
\hline
GdN& 
7.0& 
$-$& &
-8.3& 
-8.5& 
\ 9.5& 
5.8 \\
\hline
GdAs& 
7.0& 
$-$& &
-7.0& 
-8.0& 
10.7& 
6 \\
\hline
& & & &
-8.5(2)& 
-9,-4.8& 
-5& 
\\
\raisebox{1.5ex}[0pt]{ErAs}
& 
\raisebox{1.5ex}[0pt]{3.0}& 
\raisebox{1.5ex}[0pt]{6.0}& 
&
& 
& 
\ 6& 
5 \\
\hline
& & & & & &
-6& 
\\
\raisebox{1.5ex}[0pt]{YbN}
&
\raisebox{1.5ex}[0pt]{1.0}&
\raisebox{1.5ex}[0pt]{3.0}& &
-7(1)&
\raisebox{1.5ex}[0pt]{-6.5}
&
\ 4.5&
0.2 \\
\hline
\end{tabular}
\footnotetext[1]{Reference \cite{roeland}}
\footnotetext[2]{Reference \cite{rem}}
\label{tab:fdata}
\end{center}
\end{table}

% --- f subsystem ---

\fsubsystem : In all cases, stable ferromagnetic solutions were found with
the $4f$ element in the 3+ state: that is 6, 7, 11, and 13 $f$ levels are
occupied in Eu, Gd, Er, and Yb, respectively; the remainder are unoccupied.
(Antiferromagnetic solutions were also found, but FM solutions are
presented here to compare with Shubnikov--de Haas (SdH) experiments.)
Occupied $4f$ levels were always dispersionless, as Fig.\,\ref{fig:band}
shows, while unoccupied states show some dispersion, reflecting their
hybridization with the \spdsubsystem.  Gd is the only $4f$ element for
which the LDA and \GWLDA\ do not put $f$ states at \efermi \cite{Ferdi96},
because the majority 4\fup\ states are filled and minority 4\fdn\ states
are empty.  The two are separated by an exchange splitting ($U$-$J$ in
\ldau\ terminology).  For the remaining $4f$ elements either the 4\fup\ or
the 4\fdn\ level is partially filled.  \qsgw\ predicts large exchange
splittings \emph{within} this channel (controlled by different combinations
of $U$ and $J$); see e.g. the ErAs DOS in Fig.\,\ref{fig:band}.
\emph{Occupied} $f$ levels are generally in reasonable agreement with
available XPS data (see Table~\ref{tab:fdata}).  In the two Er compounds
(Er and ErAs), occupied \fup\ and \fdn\ levels are fairly well separated.
The shallower 4\fdn\ levels likely correspond to the XPS peak between
$-4.5$ and $-5$\,eV, and the 4\fup\ levels to the broad XPS peak between
$-8$ and $-10$\,eV shown in Fig.\,\ref{fig:band}.  In YbN, the separation
between occupied \fup\ and \fdn\ is small, and the XPS peak (whose width is
$\sim$3\,eV) probably corresponds to some average of them.  More precise
identification is not possible because multiplet effects are not included.
In contrast, the \emph{unoccupied} $4f$ levels are
systematically higher than observed BIS peaks, typically by $\sim$3-4
eV.  The only exception is ErAs, where the overestimate is closer to 1\,eV.
(This may well be an artifact of final-state effects in ErAs, as suggested
in Ref.~\onlinecite{komesu03}.)

Overall, the 4\fup-4\fdn\ splitting is 16.2\,eV in Gd, and $\sim$18\,eV in
GdN and GdAs.  This \emph{change} in the splitting is reflected in the
BIS-XPS data (12.2\,eV for Gd, 14\,eV for GdN and GdAs).  The carrier
concentration at \efermi\ is larger in Gd than in GdAs, which results in a
larger dielectric response, and more strongly screened $U$.

%Because the $f$ states are well removed from \efermi, these moments are
%always essentially integer
The \emph{orbital moments} folllow what is expected from Hund's
rule.  The spin moments are a little overestimated in the metals
Er and Gd, following the trend observed in $3d$ magnetic systems
such as MnAs\,\cite{Mark}.

% Reference to Walter's paper here?

Deviations from experiment can be qualitatively explained as follows.
\qsgw\ overestimates unoccupied states in $sp$ semiconductors by
$\sim$0.2\,eV\,\cite{Mark}.  The overestimate is somewhat larger in itinerant
TM oxides such as SrTiO$_3$ and TiO$_2$ ($\lesssim$1\,eV), and is larger
still ($\gtrsim$1\,eV) in the correlated oxide NiO\,\cite{faleev,qpmethod}.
This can be understood as a neglect of electron-hole interactions
(excitonic effects).  Short-range, atomic-like excitations shift peaks in
$\rm{Im}\epsilon(\omega)$ to lower frequency, increase the screening and
reduce the strength of $W$.  It is to be expected that the more localized
states are, the stronger electron-hole correlations will be ($4f>$
localized $3d>$ itinerant $3d>$ $sp$).  Indeed, when
$\rm{Im}\epsilon(\omega)$ is calculated through the Bethe-Salpeter
equation, it is in dramatically better agreement with experiment in $sp$
systems (see, e.g.  Ref.~\onlinecite{Benedict98}).  The expected
corrections to $W$ are consistent with the observed trends, both in the
overestimate of unoccupied QPEs (increasing with localization), and the
overestimate of magnetic moments.  However, is also likely that vertex
corrections to $GW$ play an increasingly greater role as localization
increases.   It is beyond the scope of our ability to include either kind of
vertex correction in \qsgw\ at present.
%Reining's paper doesn't seem to be in print anywhere.
%Indeed, in a recent \qsgw\ calculation in Cu$_2$O, peaks in the
%imaginary part of the RPA dialectric function were red-shifted when ladder
%diagrams were added\,\cite{xxx}; the resulting dielectric function was in
%very good agreement with ?xxx?.
%S. Albrecht, L. Reining, R. Del Sole, and G. Onida, Phys. Rev. Lett. 80, 4510 (1998).
%L. X. Benedict, E. L. Shirley and R. B. Bohn, Phys. Rev. B 57, R9385 (1998).
%M. Rohlfing and S. G. Louie, Phys. Rev. Lett. 81, 2312 (1998).
%F. Bechstedt, K. Seino, P. H. Hahn, and W. G. Schmidt, Phys. Rev. B 72, 245114 (2005) 
%F. Bechstedt, K. Tenelsen, B. Adolph, and R. Del Sole, Phys. Rev. Lett. 78, 1528 (1997).
% Mentions solution of Bethe-Salpeter equivalent to ladder diagrams (Mahan)

% --- spd subsystem ---

\begin{table}[b]
\caption{Gd dHvA Frequencies (T) for a magnetic field oriented along the [0001]
direction.  $\gamma1$ originates from the majority 
$6s$ band and $\alpha1$ from the majority $5d$ band.  Both lie in the
$\rm{\Gamma KM}$ plane; they are depicted in Ref.\,\cite{Mattocks}.
}
\begin{ruledtabular}
\begin{tabular}{c|cccc}
& \efermi{}$-0.2$\,eV   &   \efermi{}$-0.1$\,eV      &   \efermi  &  Expt\,\cite{Mattocks} \\
\hline
$\alpha1$       &4934  & 4260  &  3585  &    4000 \\
%\hline
%$\alpha2$       &  &   &        &  1350 \\
\hline
$\gamma1$       &7209  & 6099  & 5177 &   6900\\
\end{tabular}
\label{tab:table1}
\end{ruledtabular}
\end{table}

\emph{The} \spdsubsystem\ comprises the states at \efermi, which
control electronic transport properties.
%They are polarized by the internal magnetic field from the \fsubsystem.
Table\,\ref{tab:table1} presents two of the de Haas--van Alphen (dHvA)
frequencies observed in Gd.  By comparing them to the calculated ones as a
function of \efermi, we can determine the shift in \efermi\ required to
match the dHvA data\,\cite{Mattocks}, and thus assess the error in
those bands at \efermi.  Table\,\ref{tab:table1} shows that the \qsgw\
$\gamma1$ and $\alpha1$ should be shifted by $\sim${$-$}0.2\,eV and
$-$0.1\,eV, respectively, consistent with precision of \qsgw\ for itinerant
systems\,\cite{Mark}.  The \qsgw\ DOS at \efermi\ (1.84 states/eV$\cdot$atom)
%(1.74  states/eV$\cdot$atom without the spin-orbit (SO) interaction)
is slightly overestimated (1.57 states/eV$\cdot$atom \,\cite{wells}).

\begin{table}[htbp]
\caption{\qsgw\ cyclotron masses $m^*$, in units of the free
electron mass $m$, and frequencies $f$ (Tesla) for GdAs and ErAs.
Three bands cross \efermi\ near $\Gamma$ (see Fig.\,\ref{fig:band}):
the heavy hole (h1),
light hole (h2) and split-off hole (sh).  Ellipsoids at X have two
inequivalent axes, e$_{\rm\bar{B}C}$ and e$_{\bar{A}}$.
%These are calculated from the standard formulas $f=(e/\hbar) A$, and
%$m^*=(\hbar^{2}/2\pi)\partial{}A/\partial{}E$, where $A$ is the area of the
%extremal cross section of the Fermi surface for a given band.
Also shown are 
SdH frequencies measured for Er$_{0.68}$Sc$_{0.32}$As\,\cite{bogaerts}.
Petukhov \emph{et al.} showed that Sc doping has a modest effect on the Fermi surface,
at least within the LDA(core-like 4$f$) approximation\,\cite{petukhov-eras-gas}.
sh$\uparrow$ and sh$\downarrow$ are not distinguished in experiments.
Notation follows Ref.\,\cite{bogaerts} except that $\uparrow$ and
$\downarrow$ are exchanged.}
\begin{ruledtabular}
\begin{tabular}{lrrrrr}
                    &\multispan2 \hfil GdAs\hfil
                                    &\multispan2 \hfil ErAs \hfil
                                                    &ErScAs  \\
                    & $m^*/m$ & $f$(T)   & $m^*/m$ & $f$(T)   & $f$(T)  \\
\hline
e$_{\bar{A}\uparrow}$\ & 0.17  &  392  & 0.16  & 452   & 386   \\
e$_{\bar{A}\downarrow}$                 & 0.15  &  95   & 0.13  & 301   & 328   \\
e$_{\rm \bar{B}C\uparrow}$ & 0.51  & 1589  & 0.49  & 1317  & 1111  \\
e$_{\rm \bar{B}C\downarrow}$                 & 0.31  &  386  & 0.44  & 887   & 941   \\
\hline
h1$\uparrow$                    & 0.34  & 1575  & 0.43  & 1642  & 1273  \\
h1$\downarrow$                  & 0.40  & 1433  & 0.45  & 1368  & 1222  \\
h2$\uparrow$                    & 0.23  &  712  & 0.26  & 726   & 612   \\
h2$\downarrow$                  & 0.26  &  571  & 0.24  & 590   & 589   \\
sh$\uparrow$                    & 0.12    &  191  & 0.06   & 174   & 150   \\
sh$\downarrow$                  & 0.08   &   9   & 0.07  & 25    &       \\
\end{tabular}
\end{ruledtabular}
\label{tab:cyclotron}
\end{table}

ErAs and GdAs may be viewed as slightly negative-gap insulators
(semimetals), with an electron pocket at X compensated by a hole pocket at
$\Gamma$ (see Fig.\,\ref{fig:band}).  Several experiments address the
\spdsubsystem\ near \efermi:
%\setcounter{Alist}{0}
%\begin{list}{({\it \roman{Alist}})\,}
%{\itemsep 0mm \leftmargin 0pt \itemindent 24pt \usecounter{Alist}\addtocounter{Alist}{0}}

%\item
$(i)$
The As-$p$-like $\Gamma_{15}$ band dispersion between $\Gamma$ and X in ErAs.
As seen in Fig.\,\ref{fig:band}, the X point has states at $-$1.2\,eV and
$-$0.8\,eV, and split-off bands at $-$3.52 eV.  A single band
from $\Gamma$ to X of width $\sim$1.5\,eV was observed by photoemission
(Fig. 7 of Ref.~\onlinecite{komesu03}).

%\item
$(ii)$
SdH frequencies $f$ and cyclotron masses $m^*$ (see Table\,\ref{tab:cyclotron}).
SdH measurements in Er$_{0.68}$Sc$_{0.32}$As\,\cite{bogaerts}
agree reasonably well with \qsgw\ calculations for ErAs.
Allen et al.\,\cite{allen2} estimated $m^*$=0.17 from dc field measurements
for ErScAs,  which is consistent with the \qsgw\ values $m^*(e_{\bar{A}\uparrow})$ =0.16,
$m^*(e_{\bar{A}\downarrow})$=0.13.  Nakanishi et al.\,\cite{nakanishi}
identified two branches in the [100] direction from dHvA
measurements in GdAs: $m^*$=0.2 ($f$=246~T) for
$m^*(e_{\bar A})$, and $m^*$=0.26 ($f$=439~T) for $m^*(h_2)$.  These are in good
agreement with \qsgw\ values in Table\,\ref{tab:cyclotron}.  Koyama et
al.\,\cite{koyama} obtained $m^*$=0.48 from a broad peak in cyclotron experiments.
This may be understood as some kind of average of masses in Table\,\ref{tab:cyclotron}.

%\item
$(iii)$
The electron concentration in the pocket at X is controlled by the
(negative) gap between $\Gamma$ and X.  The \qsgw\ result agrees with experiment
to within the reliability of the calculated bandgap ($\sim$0.2\,eV):
\begin{tabular}{ll@{\quad}lll}
%{\quad} ErAs: & \ 3.5(4.0) \qsgw;  &  $3\pm{}1$  Nakanishi et al.\,\cite{nakanishi}\\
%{\quad} GdAs: & \ 3.3(3.6) \qsgw;  &  2.3        Allen et al.\,\cite{allen}
\quad  ErAs: & \ 3.5$\times$10$^{20}$~cm$^{-3}$ \qsgw;  &  $3\pm{}1$& expt, Ref\,\cite{nakanishi}\\
\quad  GdAs: & \ 3.3$\times$10$^{20}$~cm$^{-3}$ \qsgw;  &  2.3      & expt, Ref\,\cite{allen}
% {\quad} ErAs: & \ 3.5$\times$10$^{20}$~cm$^{-3}$ \qsgw;  &  $3\pm{}1$  Nakanishi et al.\,\cite{nakanishi}\\
\end{tabular}
\\
%Data in parentheses are \scld\ results.  The concentration is controlled by the
%(negative) gap between $\Gamma$ and X.  This level of agreement is consistent
%with the reliability of the calculated bandgap ($\sim$0.2\,eV).

%\end{list}

\begin{figure}[htbp]
\includegraphics[angle=0,width=0.3\textwidth,clip]{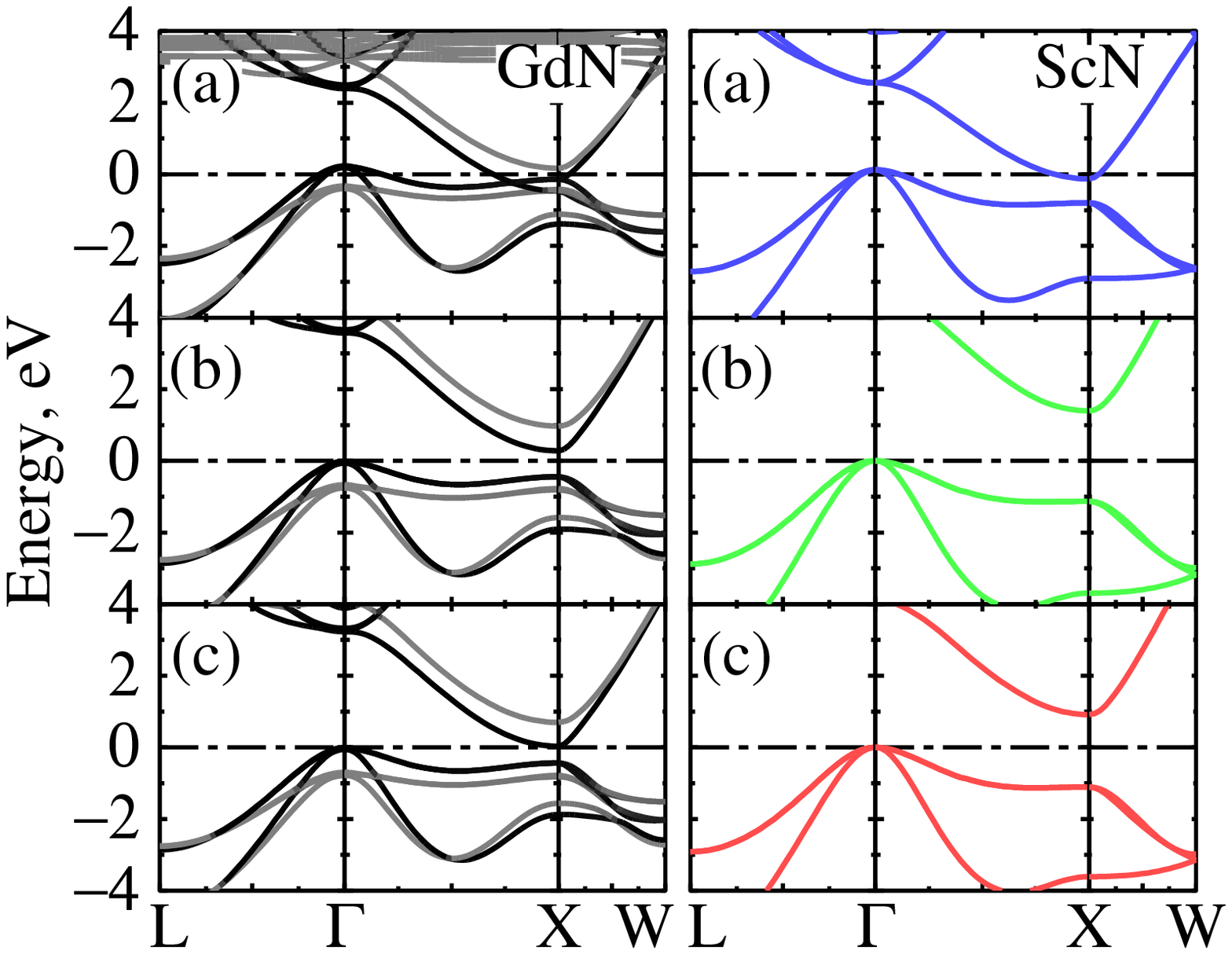}
\includegraphics[angle=0,width=0.1\textwidth,height=0.23\textwidth,clip]{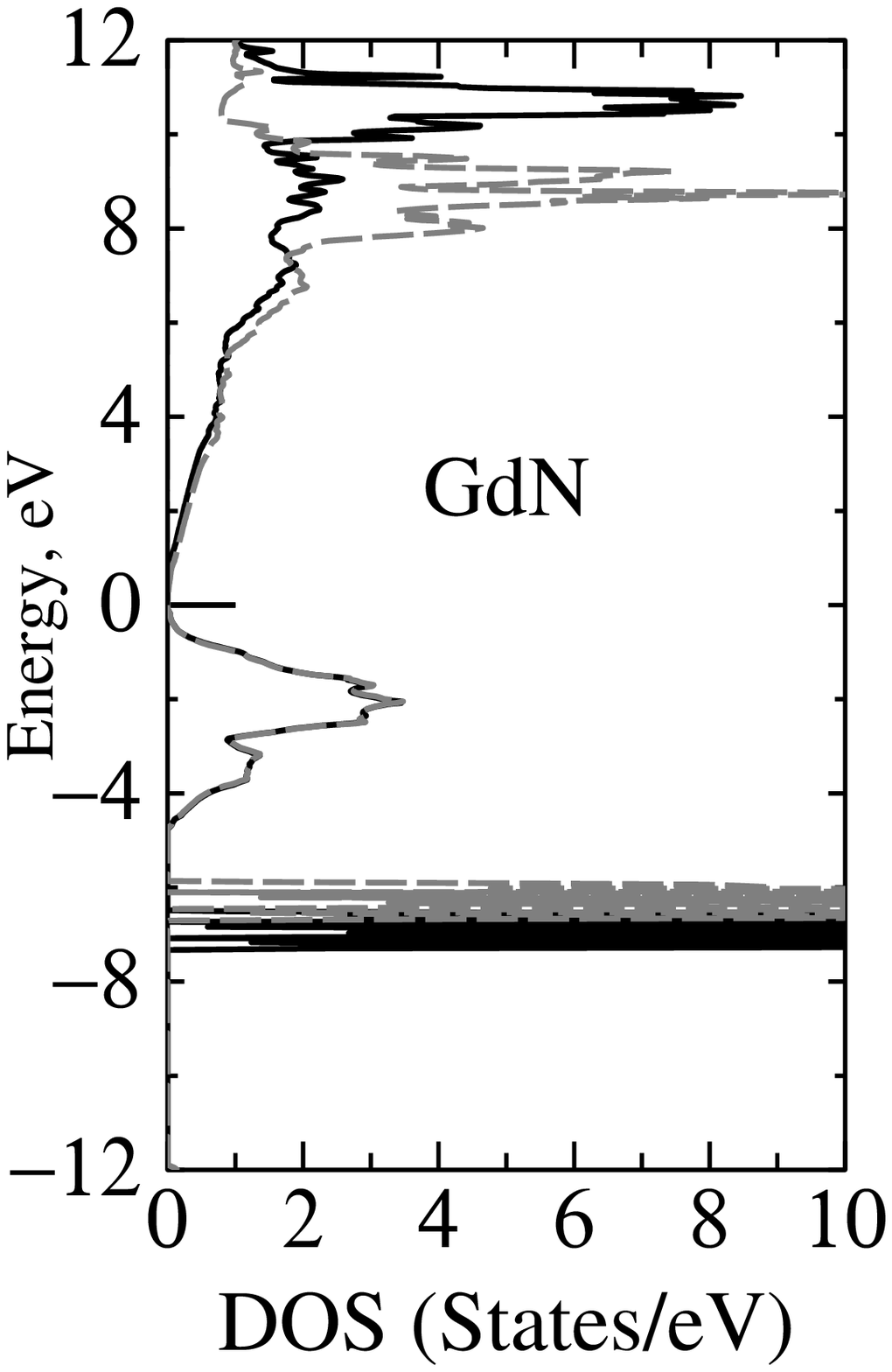} % The DOS and scaled DOS
\caption{
Energy bands near \efermi\ (\efermi=0) in GdN and ScN.
For GdN, majority (minority) bands are drawn in dark (light) 
lines. $(a)$ \ldau\ for GdN ($U$=6.7\,eV, $J$=0.69\,eV) and LDA for ScN.
A semimetal is predicted in both cases.
%In both cases, the conduction band at X falls slightly
%below the valence band at $\Gamma$,
%resulting in an electron pocket at X and a hole pocket at $\Gamma$.
$(b)$ \qsgw\ bands:  both materials have positive gaps.
$(c)$ `scaled $\Sigma$' bands (see text).
The ScN $\Gamma$-$\Gamma$ and $\Gamma$-X
gaps fall close to measured values
($E_c({\rm X})$-$E_v(\Gamma)${}=0.9$\pm$0.1\,eV;
$E_c({\rm X})$-$E_v({\rm X})${}=2.15\,eV\,\cite{Hamad}).
%The GdN gap is predicted to be $\sim$0.05\,eV.
Right panel: DOS in \qsgw\ (dark lines) and `scaled $\Sigma$' (light lines).
Scaling reduces the 4\fdn-4\fup\ splitting, but
the \spdsubsystem\ is little affected by it.
%because the $f$-levels are still far removed from \efermi.
}
\label{fig:bandn-scn}
\end{figure}

GdN is qualitatively similar to GdAs and ErAs, but
there is some confusion as to whether GdN is an insulator or semimetal.
Wacher and Kaldis measured a large carrier concentration (1.9$\times$10$^{21}$\,cm$^{3}$).
%While the observed optical absorption edge of $\sim$1\,eV\,\cite{kaldis1}
%suggests it is a semiconductor, the large carrier concentration
%(1.9$\times$10$^{21}$\,cm$^{3}$) measured by Wacher and Kaldis
%\,\cite{kaldis2} indicates that it is a semimetal. 
But, direct measurements of resistivity indicate insulating
behavior\,\cite{xiao}.  Remarkably, \qsgw\ predicts \emph{two kinds} of
stable, self-consistent solutions near the observed lattice constant
$a$=5.00: one is insulating with $E_g(a)$=$E_c({\rm
X})$-$E_v(\Gamma)${}={}$+0.2$\,eV, and
$\partial{}E_g/\partial{a}{}={}2.7$\,eV/\AA. It is stable for
$a${}$>$4.96\AA.  The other is a semimetal with $E_g(a)${}={}$-0.2$\,eV
and $\partial{}E_g/\partial{a}={}2.1$\,eV/\AA.  It is stable for
$a${}$<$4.99\AA\ and at larger lattice constants when a tetragonal or
trigonal shear is applied.  Thus, a range of structures is found for which
two solutions coexist, and the MIT is \emph{first-order}.  It can be
connected with discontinuous changes in the dielectric function:
$\epsilon$($\omega${}$\to$0) diverges in the semimetallic phase, and
approaches a constant in the insulating phase.

%Except for spin polarization, the electronic
%structure around \efermi\ is similar to ScN.
%LDA predicts the $\Gamma-$X gap to be negative in ScN,
%though is  measured to be $E_g$=0.9$\pm$0.1\,eV \cite{Hamad}; see
%Fig.\,\ref{fig:bandn-scn} (a).  (The direct X-X gap is 2.15\,eV).
%A similarly negative gap is predicted for GdN by \ldau;
%\ldau\ is essentially LDA-like for the $spd$ subsystem.
%\qsgw\ predicts a positive bandgap in both GdN
%and ScN as panel $(b)$ shows, with $E_g${}$\sim$0.22\,eV in GdN.

To correct for \qsgw's tendency to overestimate gaps slightly, we adopt
the `scaled $\Sigma$' approach\,\cite{chantis}: we take a linear combination
of the LDA and the \qsgw\ potentials, $(1-\alpha) \times$ \qsgw + $\alpha
\times$ LDA.  We found that $\alpha$=0.2 can accurately reproduce
experimental band gaps for a wide variety of materials; see
Ref.~\onlinecite{chantis} for III-V and II-VI semiconductors. Energy bands
with $\alpha$=0.2 are shown in Fig.\,\ref{fig:bandn-scn}(c).  For
comparison, we also show the bands of ScN, whose electronic structure is
similar to GdN near \efermi.  Scaling brings the ScN $\Gamma$-X and X-X
gaps to within $\sim$0.1\,eV of experiment, consistent with our general
experience.  It is reasonable to expect that GdN will be similarly
accurate.  Panel $(c)$ shows that for GdN, $E_g${}$\approx$0.05\,eV.  Thus GdN is
right on the cusp of a MIT.  The spin-averaged X-X gap (1.48\,eV for
majority, and 0.46\,eV for minority) is in close agreement with 0.98\,eV
measured in paramagnetic GdN\,\cite{kaldis1}.

%Remarkably, \qsgw\ predicts MIT to be \emph{first-order}.
%We obtain two stable solutions: one is metallic with
%$E_g${}$\sim${}$-$0.3\,eV; the other is insulating as described above.
%The first-order MIT may be caused by slight external perturbations such as
%shear strain typical for epitaxial growth on mismatched substrates,
%alloying, and so on.  It occurs as a
%cooperative phenomenon connected with the rapid increase of the dielectric
%response as the bandgap closes.  (\qsgw\ predicts a similar MIT for fcc
%YH$_3$\cite{sakuma_yh3}).  This theoretical prediction of a first-order MIT
%may account for the contradictory experimental data in GdN mentioned in the
%introduction.

% --- compare to \ldau ---
{\it Comparison between \qsgw\ and \ldau}:
In both methods, the one-particle effective potential
is written as $V^{\rm{eff}} = V^{\rm{LDA}} + \Delta V$.
In \qsgw, ${\it \Delta}V$ is rather general;
in \ldau, ${\it \Delta}V$ is added only for the onsite $\ff$ block
which is specified by $U$ and $J$\cite{anisimov}.
For comparison, we extract
the onsite $\ff$ block (\qsgwff) from $V^{\rm{eff}}$
given by \qsgw, by Fourier transform techniques\,\cite{qpmethod}.
%It is done using a Fourier transform to map
%${\it \Delta} V^{\rm{eff}}$ to real space\cite{qpmethod}.
The \spdsubsystem\ near \efermi\ almost perfectly recovers the LDA-like
bands of the LDA+$U$ method, Fig.\,\ref{fig:bandn-scn}(a); the \fsubsystem\
remains essentially unchanged from the full \qsgw\ calculation,
Fig.\,\ref{fig:band}(c).  Thus, the \qsgw\ $H_0$ may be thought of as the ``ultimate''
LDA+$U$ Hamiltonian, with Hubbard parameters connecting all orbitals between all
sites in a basis-independent way.  Significantly, \emph{no choice} of the
standard \ldau\ parameters can reproduce the \qsgw\ (\fdn,\fup) levels
simultaneously, because their average position is essentially determined
from the LDA.

\begin{acknowledgments}
This work was supported by ONR contract N00014-02-1-1025.
We are also indebted to the Ira A. Fulton High Performance Computing
Initiative.  A. N. Chantis would like to thank A. G. Petukhov
for fruitful discussions.
\end{acknowledgments}

\bibliography{fshell}
\end{document}